\documentclass[fleqn,usenatbib,useAMS]{mnras}

\usepackage{graphicx}	% Including figure files
\usepackage{amsmath}	% Advanced maths commands
\usepackage{amssymb}	% Extra maths symbols
\usepackage{multicol}        % Multi-column entries in Tables
\usepackage{bm}		% Bold maths symbols, including upright Greek
\usepackage{pdflscape}	% Landscape pages
\usepackage{graphicx}	% Including figure files
\usepackage{natbib}
\usepackage{hyperref}
\title[Circularly polarised emission from exoplanets]{A search for circularly polarised emission from young exoplanets }

\author[C. R. Lynch et al.]{
C. R. Lynch,$^{1,2}$\thanks{E-mail: clynch@physics.usyd.edu.au}
Tara Murphy,$^{1,2}$
D. L. Kaplan,$^{3}$ 
M. Ireland,$^{4}$ 
M. E. Bell$^{2,5}$
\\
$^1$ Sydney Institute for Astronomy, School of Physics, The University of Sydney, NSW 2006, Australia\\
$^2$ ARC Centre of Excellence for All-sky Astrophysics (CAASTRO)\\
$^3$ Department of Physics, University of Wisconsin--Milwaukee, Milwaukee, WI 53201, USA\\
$^4$ Research School for Astronomy $\&$ Astrophysics, Australia National University, Canberra, ACT 2611, Australia\\
$^5$ CSIRO Astronomy and Space Science, PO Box 76, Epping, NSW 1710, Australia
}

\date{Accepted XXX. Received YYY; in original form ZZZ}

\pubyear{2016}

\begin{document}
\label{firstpage}
\pagerange{\pageref{firstpage}--\pageref{lastpage}}
\maketitle

\begin{abstract}
We report the results of a 154 MHz survey to search for emission from exoplanets located in the Upper Scorpius subgroup of the Sco Cen OB2 Association, the closest substantial region of recent star formation. This survey is different from previous efforts in that it is the first to target exoplanets orbiting Myr-old stars. Young exoplanet systems are expected to be the best candidates for radio detections given the higher magnetic field strengths predicted for young planets as well as the stronger and more dense stellar wind expected for the host stars. The radio emission from exoplanets is expected to be highly circularly polarised therefore we restricted our search to the circular polarisation images rather than the total intensity images. We carried out two different search methods using this data. The first method was a targeted search for exoplanet emission using catalogues of known stars and Hot Jupiters within the Upper Scorpius field. The second search method was a blind search for highly circularly polarised sources in the field and for sources identified only in our polarisation images. Both the blind and targeted search resulted in non-detections with typical 3$\sigma$ flux density limits of 4--235\,mJy over timescales of 1.87--1000\,minutes. In particular, we place the first limits on low-frequency emission from the Hot Jupiter systems WASP-17 b and K2-33 b. These are the first results from a larger program to systematically search for low-frequency radio emission from planets orbiting young stars. 

\end{abstract}


%%%%%%%%%%%%%%%%%%%%%%%%%%%%%%%%%%%%%%%%%%%%%%%%%%

%%%%%%%%%%%%%%%%% BODY OF PAPER %%%%%%%%%%%%%%%%%%

\section{Introduction}

Planets capable of generating strong, planetary-scale magnetic fields are expected to produce bright radio emission \citep{Winglee:1986, Zarka:2001}. The magnetised planets in our own Solar System are known to emit intense, low-frequency radio emission from their auroral regions through the electron-cyclotron maser instability (CMI). This emission arises from the propagation of energetic (keV) electrons along converging magnetic field lines in the planet's magnetosphere. The source of the electron acceleration is an interaction between the solar wind and the planet's magnetosphere. CMI emission is bright, highly beamed, 100$\%$ circularly polarised, and sporadic with variability on time-scales ranging from seconds to days \citep{Wu:1979, Treumann:2006}. Analogously, magnetised exoplanets are expected to be sources of bright, low-frequency radio emission \citep[e.g.][]{Winglee:1986, Zarka:2001}. 

There are over 3000 confirmed exoplanets \citep{Schneider:2011}, most of which were discovered through indirect means involving a search for the exoplanet's influence on its host star. Due to the large contrast in brightness between the exoplanet and the host star, far fewer of the known exoplanets have been discovered through direct imaging \citep{Perryman:2011}. In contrast, radio emission from magnetised exoplanets is expected to exceed the emission of the planetary host star \citep{Griessmeier:2005} and radio observations could provide another method of direct detection of exoplanets. Radio observations would also provide a direct measurement of the planet's surface magnetic field strength and in turn provide insight into the interior composition of these planets. Furthermore, \citet{Hess:2011} show that the variability of the radio emission in both time and frequency can provide constraints on the rotational period of the planet, the orbital period and inclination, and the magnetic field tilt relative to the rotation axis.

Theoretical estimates of the main characteristics of planetary radio emission can be used to select the best targets for observations. Both the expected radio flux density and the emission frequency need to be considered when choosing potential targets. The `Radiometric Bode's law' is an empirical relation, based on observations of the magnetised Solar System planets, that is primarily used to predict the intensity of radio emission from exoplanet systems. This relation relates the incident Poynting or kinetic energy flux of the Solar wind to the radio power produced by a planet \citep{Desch:1984,Zarka:2001}. Note that this relation does not mean that auroral radio emission produced by all planets is primarily driven by the stellar wind. For example, the auroral emission observed from Jupiter is mainly driven by its rotation and not the Solar Wind \citep{dePater:2015}.

In the last few years there have been a number of theoretical studies that have estimated the expected radio flux densities from exoplanet systems using the Radiometric Bode's law. \citet{Farrell:1999} focused on five Solar-like stars with Jupiter systems, finding $\tau$ Bootes to be an optimal target. \citet{Lazio:2004} expanded on this work and modelled the radio flux density for over 100 sources, noting that planets with small orbital distances should produce mJy level emission. Further work by \citet{Stevens:2005} and \citet{Griessmeier:2005} focused on investigating the effect the stellar wind and mass loss rate have on the expected flux density. The results from their modelling find that, in the case of kinetic energy driven Radiometric Bode's law, objects with higher mass-loss rates and wind velocities relative to the Sun are more favourable targets for detections of radio emission. 

\citet{Griessmeier:2007} compared the kinetic and magnetic energy models for a planet-star interaction. They argued that while the magnetic energy Radiometric Bode's law produces a higher radio flux density, both energy input models need to be considered since it is not clear which dominates in planet-star systems. This idea is supported by the work of \citet{Zarka:2007} who found that for the Solar System planets it is not possible to determine which incident power, magnetic or kinetic, drives the CMI emission. 

Many of these authors note that precise modelling of the stellar wind is crucial for estimating accurate radio flux densities. This is because the emitted radio power predicted using the Radiometric Bode's law is heavily dependent on the stellar wind parameters. Three dimensional numerical simulations of the stellar winds of both Solar-like stars \citep{Vidotto:2015} and pre-main sequence stars \citep{Vidotto:2010} have been used to estimate the emitted radio power from Hot Jupiter systems. These efforts suggest that Hot Jupiters orbiting young stars are the best candidates for radio detections if the Radiometric Bode's law holds for all planets. 

Counter to the above efforts, \citet{Nichols:2011, Nichols:2012} suggest that radio emission from some exoplanets may not be dominated by the star-planet interaction and therefore the intensity of the expected radio emission would not follow the Radiometric Bode's law. Rather these planets may generate their auroral emission in a similar manner to Jupiter, through a magnetosphere-ionosphere coupling current system associated with an internal plasma source such as an active moon.  In this case, fast-rotating massive planets orbiting in large orbits around XUV-bright stars are capable of generating detectable radio emission. Further arguments against the Radiometric Bode's law suggest that for close-in planets magnetospheric convection will saturate. These systems will not be able to dissipate the total incident stellar wind energy and thus the Radiometric Bode's law will over estimate the flux densities substantially in these cases.

CMI emission occurs near the cyclotron frequency of the emitting electrons, 
\begin{equation}
\nu_{c} = \frac{e B_p}{2\pi m_e} = 2.8\ B_p\ \text{MHz}
\end{equation}
where $m_e$ and $e$ are the electron mass and charge, and $B_p$ is the local polar planetary magnetic field strength in units of Gauss \citep{Farrell:1999}. Note for a dipolar magnetic field, $B_p$ is equivalent to twice the equatorial field at a fixed radial distance. Thus planetary CMI emission is expected to have a high frequency cut-off related to regions of maximum magnetic field strength near the surface of the planet. This means that  predicting the magnetic field strength of potential exoplanet targets is important for  determining which can be observed with current low frequency telescopes. There are currently two different theoretical approaches for determining the magnetic field of an exoplanet. \citet{Griessmeier:2004} and \citet{Farrell:1999} use the balance between the Coriolis force, Lorentz force, buoyancy and pressure gradient forces to calculate the planetary magnetic moment and find that it strongly depends on the rotation rate of the planet. The second method of estimating a planet's magnetic field relates the amount of energy flux available from the planet's core to the  magnetic field strength. This method finds that the field strength has no dependence on the rotation of the planet and favours young Hot Jupiters \citep{Christensen:2009}. 

The first non-targeted searches for radio emission from exoplanets occurred before the first detection of the currently known population of exoplanets \citep{Winglee:1986}. More recently, searches have involved targeting nearby Hot Jupiters previously detected through radial velocity and transiting observations. Despite the numerous searches for radio emission from exoplanets there have been no unambiguous detections to date \citep[e.g.][]{ Bastian:2000, Lazio:2004, Lazio:2010, George:2007, Smith:2009, Stroe:2012, Hallinan:2013, Lecavelier:2013, Sirothia:2014, Murphy:2015}. These attempts rely on the results of optical techniques that tend to be biased towards exoplanet systems around main sequence stars \citep{Crockett:2012,Lagrange:2013, Johns-Krull:2016}. However, from modelling results using the Radiometric Bode's law, its possible that the best targets are Hot Jupiters around pre-main sequence stars since their stellar winds and mass loss rates are much greater \citep{Griessmeier:2005,Vidotto:2010}. 

The Upper Scorpius Association is a region of young stars with ages ranging from $10-20$ Myr and and is an ideal region to look for radio emission from young exoplanets. This is the closest region with recent star formation, located at a distance of 145 pc, and is thought to contain a few thousand forming stars \citep{deZeeuw:1999}. It is unknown what fraction of forming stars host Hot Jupiter systems, but for Solar-type main sequence stars about 1$\%$ host Hot Jupiter systems \citep{Wright:2012}. So we naively expect $\sim$10 Hot Jupiters in the Upper Scorpius Association, if migration of Hot Jupiters is contemporaneous with gas disk dispersal. Some recent work suggests that Hot Jupiters may migrate to their close orbits after time periods of order 100 Myr, due to planet-planet scattering or Kozai-Lindov resonances \citep[e.g.][]{Knutson:2014}. Some giant planets certainly migrate at very young ages \citep{Mann:2016}, but the lack of a population of transiting higher-mass giant planets hints that a significant fraction of Hot Jupiters may migrate significantly after gaseous disk dispersal.  A dominant late migration scenario for Hot Jupiter formation would produce a paucity of radio-bright young Hot Jupiters. We targeted the Upper Scorpius Association with the Murchison Widefield Array \citep[MWA;][]{Tingay:2013}. The MWA has a field-of-view of 144 square degrees which allowed us to do a blind search of several hundred objects simultaneously. This is the first radio survey to target young exoplanets around forming stars. By targeting a region of best candidates rather than relying on previous detections in other wavebands we can avoid any biases or selection effects in those wavebands.  

\section{Observations and Data Reduction}
We  observed the Upper Scorpius Association at 154 MHz, averaging over the full 30.72 MHz bandwidth of the MWA. The observations were carried out between 2015 June 01 -- 05. We observed for between 7.3 and 7.6 hours per night, for a total of 37.82 hours. These observations were recorded as a set of scans with a duration of 1.87 minute (or 1.87 minutes). Each observation scan was reduced and imaged as outlined below. 

\begin{table}
 \centering
  \caption{Measured 3$\sigma$ noise limits for images over different timescales}
  \label{tab:NoiseEst}
  \resizebox{\columnwidth}{!}{\begin{tabular}{lcccc}
    \hline
   Image & Time  & Stokes I & Absolute Stokes V & Stokes V\\
     & (minute) & (mJy) & (mJy) & (mJy) \\
    \hline
    Snap-shot & 	1.87 	& 235.	& 110. 	& 110.  \\
    Hourly & 		60. 	& 68. 	& 37.	& 20.	\\
    Nightly & 		200. 	& 46. 	& 31. 	& 12.	\\
    Full Data-set & 1000. 	& 39.  	& 30.	& 4. 	\\
    \hline
  \end{tabular}}
 \end{table}

To calibrate the set of observations for each night we additionally observed 3C444, a bright calibration source with well-modelled emission, for 1.87 minutes each night. This source was used to produce a time-independent, frequency dependent amplitude and phase calibration solution that was applied to the rest of that nights two minute scans. For each scan the visibilities were pre-processed using the \textsc{cotter} MWA preprocessing pipeline. This pipeline performs the following tasks: flagging radio-frequency interference  using the \textsc{aoflagger}  \citep{Offringa:2012}, averages the data, and converts the  data into a \textsc{casa} measurement set. Each of the 1.87 minute scans  was then imaged using the \textsc{wsclean} algorithm  \citep{Offringa:2014}. For the imaging  a pixel size of 0.75 arcsec and image size of 3072 pixels was used.  
For each 1.87 minute scan we produced a total intensity, Stokes I, image with  Briggs weighting of $-1$  (closer to uniform weighting) and a circular polarisation, Stokes V, image with Briggs weighting of  $+1$ (closer to natural weighting).  The  \textsc{wsclean} algorithm produces these two images by forming a $2 \times 2$ complex Jones matrix  $I$ for each image pixel.  A primary beam correction was performed by inverting the beam voltage matrix $B$ and computing $B^{-1}IB^{*-1}$, where $*$ denotes the conjugate transpose (for further details see \citet{Offringa:2014}). 

Analysis of the full set of 1.87 minute images across all five nights of observations showed that the flux densities for the sources in both Stokes I and V images were severely affected at low elevations. This is due to the foreshortening of the projected baselines at low elevations resulting in beam elongation in the N-S direction. To avoid false detections in our Stokes V images at low elevations we removed  the first and last 50 images for each night of observation. This left us with 100 images for a total of 3.3 hours per night. For these images we checked the flux calibration with that of the recently published GaLactic and Extragalactic All-sky MWA survey \citep{Hurley-Walker:2017}. We find that the flux density scaling is within a factor of two of this survey, which is acceptable for the following analysis. 
 
\subsection{Median Images}
The expected flux density of an exoplanet source is less than a few tens of mJy and so to achieve the required sensitivities we median stacked our 1.87 minute images to create images covering longer integrations. We did not need to regrid the 1.87 minute images as they are already on a common coordinate grid, centred at an RA and Dec. of 240.0, $-30.0$ degrees. We chose to use the median of the pixel values rather than the mean because the mean is more sensitive to poor quality images. 

Before stacking our images we needed to take into account the variability time scale expected for planetary emission. Due to the narrow beaming of the expected CMI emission, radio emitting planets will only be detected over ranges of rotational and orbital phases during which their active magnetic field lines are suitably oriented relative to our line of sight. Further, \citet{Hess:2011} simulated the expected dynamic spectra for Hot Jupiter systems and showed that the detectable emission covers only a few percent of the rotational or orbital phase. Because the rotational period of exoplanets are unknown and exoplanets orbiting close ($< 0.1$ AU) to the host star are expected to be tidally locked \citep{Griessmeier:2004}, we assume the rotational period of the potentially detected Hot Jupiters to be the same as their orbital period. The majority of the currently known Hot Jupiter systems have orbital periods between $1 - 6$ days \citep{Schneider:2011}. From the simulated dynamic spectra of \citet{Hess:2011}, we assume the emission lasts up to 5\% of the orbital phase and consequently we expect to see variability on time-scales of $1 - 7$ hours. To take into account this variability we median-stacked the 1.87 minute images to create images on two time-scales: hourly and nightly. We also stacked all 500  of the remaining images to produce a single deep image and place the best upper limits on the expected radio flux density in the case of non-detections.

Furthermore we need to also consider the possibility of polarisation reversal. CMI emission can either be left or right handed circularly polarised depending on the hemisphere from which it is emitted \citep{Zarka:1998, Lamy:2008}. Thus the polarisation of the emission observed with the MWA will be dependent on which hemisphere is visible to the telescope and can change with either the planet rotation or orbit, as shown by \citet{Hess:2011}. Polarisation helicity reversal has been observed in other sources of CMI emission
\citep[e.g. ultracool Dwarfs:][]{Hallinan:2007, Lynch:2015} and should be considered when stacking Stokes V images. To avoid de-polarising our images through stacking over opposite helicities, we two sets of Stokes V median stack images, one in which we take the absolute value of all the pixels before creating the median stack images in Stokes V.  

The average sensitivity achieved in each of the stacked images is listed in Table \ref{tab:NoiseEst}, including both cases of Stokes V stacking (absolute values and not). To estimate the noise in each image we  use the 99.7 percentile of the pixel distributions (equivalent to a 3$\sigma$ limit). Due to the compactness of the MWA, it has been shown that the noise in the Stokes I images will be dominated by classical and side-lobe confusion \citep{Franzen:2015, Rowlinson:2016}. The Stokes I rms values we measure are consistent with these previous studies. Due to the much smaller number of sources with circular polarisation, our Stokes V images are not confusion limited. The noise in these images is expected to dominated by thermal noise of the telescope. However stacking the absolute value of the pixels causes the final stacked Stokes V images to be somewhat less sensitive than expected. Comparing to the RMS noise for median stacked images without using the absolute pixel values, the noise is increased in our images by a factor of 1.85 for the 60 minute images, 2.58 for 200 minute images, and 7.5 for the 1000 minute image.  The median stack of the absolute pixel-values is clearly not the optimal statistic to use  for combining these images and further work is needed to determine what is the best way to combine images to increase the sensitivity without reducing the polarised signal.

\section{Searching for Exoplanets}

For each set of images (hourly, daily, and full observation) we used the source finder \textsc{aegean} \citep{Hancock:2012} with a 5$\sigma$ detection threshold to create catalogues of significant sources within the Stokes I and V images. Due to the imperfect MWA primary beam model, the flux density measurements far from the phase centre of the observation can have errors of up to 10$\%$ \citep{Hurley-Walker:2014}; with this in mind we restrict our search to sources located within the primary beam of the image or within a distance of 12 degrees from the phase centre.  These three sets of catalogues were then used to carry out two different methods to search for exoplanet emission in the images. Since CMI emission from  planetary systems is known to be  highly circularly polarised \citep{Wu:1979, Treumann:2006}, in this analysis we focus solely on  the circularly polarised sources. This has the added benefit that our Stokes V images are more sensitive than our Stokes I images (see Table \ref{tab:NoiseEst}).

\subsection{Targeting known sources within the field}
We carried out a targeted search for exoplanet emission  using catalogues of known stars  and exoplanets within the Upper Scorpius field. For each of the catalogues we first cross-matched the catalogue with our two catalogues (absolute and normally stacked) of Stokes V sources in the hourly, daily and full median stack images. Using the results of that cross-match we then carry out a second cross-match with the Stokes I catalogues and calculate the fractional polarisation for sources common to both Stokes catalogues. We then investigate those sources with fractional polarisation greater than 0.5. Because our Stokes I images are less sensitive that our Stokes V images, we also investigate sources which only have matches in our Stokes V catalogues. 

\begin{table}
\centering
  \caption{Physical Parameters for Hot Jupiters within the Upper Scorpius field as defined in section \ref{sec:planets}}
  \label{tab:HotJup}
  \begin{tabular}{lcc}
    \hline
    & K2-33 b & WASP-17 b\\
    \hline
    $d$ (pc) & 145 & 400\\
    $M_p$ (M$_{\text{J}}$) & 3.6 & 0.486\\
    $R_p$ (R$_{\text{J}}$)& 0.45 & 1.991\\
    $a$ (AU) & 0.041 & 0.0515\\
    $t$ (Gyr) & 0.0095 & 3.0 \\ 
    $\dot{M}$ (M$_{\odot}$ yr$^{-1}$) & 10$^{-9}$ & 4$\times$10$^{-14}$\\
    B$_{p,\text{G}}$ (G) & [0.8 ... 25.0]$^{\text{a}}$ & [0.008 ...0.03]$^{\text{a}}$\\
    $\nu_{c,\text{G}}$ (MHz) & [2.2 ... 70.0] & [0.02 ...0.09]\\
    $S_{\nu_{c,\text{G}}}$ (mJy) & [11.7 ... 37.0] & [0.16 ...0.26]\\
    B$_{p,\text{R}}$ (G) & 124$^{\text{b}}$ & 6$^{\text{b}}$\\
    $\nu_{c,\text{R}}$ (MHz) & 346 &  16  \\
    $S_{\nu_{c,\text{R}}}$ (mJy) & 34.0 & 0.01 \\
    $S_{154}$ (mJy)&  9.7 & 8.4 \\    
    \hline  
  \end{tabular}\\
 $^{\text{a}}$ Using the scaling laws in \citet{Griessmeier:2004}\\
 $^{\text{b}}$ Using the scaling laws in \citet{Reiners:2009}
 \end{table}
 
While the Sco Cen OB2 Association is nearby, the identification of stars within this association is not complete for most mass regimes. The exception is for the most massive stars, whose membership has been fully characterised by \citet{Rizzuto:2011}. They identify 436 high mass members located within Sco Cen OB2. There are expected to be $\sim$2000 pre-main sequence late K- and M-type stars located within the Upper Scorpius \citep{Preibisch:2002}. \citet{Rizzuto:2015} have started to characterise this population of stars and identify 237 spectroscopically confirmed stars in Upper Scorpius.  Combining these two catalogues we have the most complete known census of the stellar population within Upper Scorpius. We use the location of these objects in our targeted search as potential hosts for exoplanet systems. Additionally, we use the Extrasolar Planetary Encyclopaedia \citep{Schneider:2011} to identify two  Hot Jupiter systems located within the field: K2-33 b \citep{Mann:2016,David:2016} and WASP-17 b \citep{Anderson:2010}. We list the relevant physical parameters measured for these two systems in Table~\ref{tab:HotJup}, including the distance to the system, $d$, the mass, $M_p$, and radius, $R_p$ of the planet, the semi-major axis of the planet's orbit, $a$, and the age, $t$, of the host star.

\begin{figure*}
\centering
 \includegraphics[width=\linewidth]{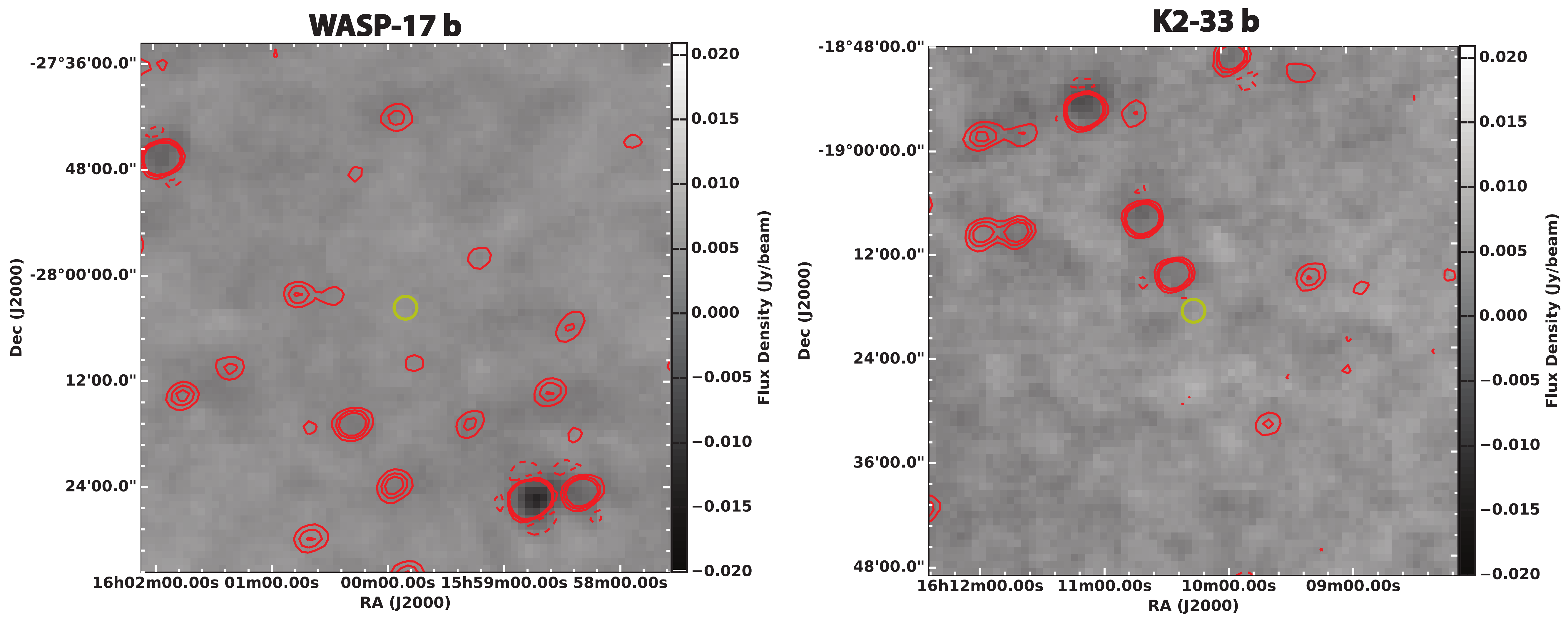}
 \caption{The 154 MHz Stokes V images overlaid with Stokes I contours for the two known Hot Jupiter systems in this field: WASP-17 b (left) and K2-33 b (right); the yellow circle marks the location of each source in their respective image. Both the contours and images are the result of the 16.6 hour median stack. The contour levels are -1, 1, 2, 3 times 39.0 mJy, the Stokes I 3$\sigma$ noise value.}
 \label{fig:planets}
\end{figure*}

\subsubsection{Hot Jupiter Systems K2-33 b and WASP-17 b}\label{sec:planets}
The two known Hot Jupiter systems were not associated with any of the Stokes V sources in our catalogues over any of the time-scales. The best limit on the Stokes V emission comes from the normally stacked Stokes V image of the full data set. The 3$\sigma$ upper limits to the 154 MHz emission are listed in Table~\ref{tab:HotJup} as $S_{154}$. We estimated this value by measuring the 99.7 percentile of the pixel values within an area several times the resolving beam centred on the location of each source. In this table we also give the predicted mass-loss rate and magnetic field strengths for both sources, which we use to calculate the expected emission frequency and radio flux density. 

For young objects like K2-33 the mass-loss rates are not well constrained by measurements and to estimate of the mass loss-rate for this object we rely on the modelling efforts of \citet{Vidotto:2010}. \citet{Vidotto:2010} use 3D numerical magnetohydrodynamic simulations to study the stellar winds of Myr-old stars that have dissipated their accretion disks. They estimate a mass-loss rate of  10$^{-9}$ M$_{\odot}$ yr$^{-1}$ for these young stars. For stars with ages greater than 700~Myr, the mass-loss rates are well modelled by the empirically derived relation $\dot{M} \propto t^{-2.00\ \pm 0.52}$, where $t$ is the age of the star \citep{Wood:2002, Lammer:2004}. We use this expression to estimate the mass-loss rate for WASP-17. The estimated mass-loss rates, $\dot{M}$, are listed in Table ~\ref{tab:HotJup}.

Depending on the physical parameters of the planet, the two theoretical models developed for predicting magnetic field strengths can differ by factors of $2 - 5$. To get a sense of what range of magnetic field strengths are predicted for these two objects, we calculate the strength of the magnetic field using both methods. \citet{Griessmeier:2004} list four different scaling relations for the magnetic moment of a planet which depend on the mass, radius, and rotation rate of the planet. Assuming WASP-17 b and K2-33 b are tidally locked to their host star and following exactly the procedure outlined in \citet{Griessmeier:2004}, we calculate the range of magnetic field strengths B$_{p,G}$ in Table \ref{tab:HotJup}. We use the scaling law by \citet{Reiners:2009} to calculate a second set of magnetic field strengths. This expression depends on the mass, radius, and luminosity of the planet. Both the mass and radius for each of these planets is already known and so we use the evolutionary tracks calculated by \citet{Burrows:1993,Burrows:1997} to estimate the luminosities for each planet. Using these values we calculate the magnetic field strengths B$_{p,R}$ in Table \ref{tab:HotJup}. We also list the associated maximum cyclotron frequency, $\nu_{c}$, that we expect given these field strengths.

Using the different sets of field strengths and mass-loss rates, we can use the expressions derived by \citet{Griessmeier:2005} for the predicted radio flux density emitted by a planet: 
\begin{equation}
\label{eq:griess}
S_{\nu} \propto R^3_p\ \mu^{-1/3}\ n_0^{2/3}\ v^{7/3} d^{-4/3}
\end{equation} This expression assumes the case of kinetic energy dominate Radiometric Bode's law. We first re-write equation \ref{eq:griess} in terms of the stellar mass-loss rate, $n_0 \propto\ \dot{M}/v$, and planetary magnetic field,\,$B \propto\ \mu/R^3_p$:

\begin{equation}
S_{\nu} \propto R^2_p\ B_p^{-1/3}\ \dot{M}^{2/3}\ v^{5/3} a^{-4/3} d^{-2}
\end{equation} where $\mu$ is the magnetic moment for a planet with magnetic field strength $B_p$, radius $R_p$ and semi-major axis $a$, $\dot{M}$ is the mass-loss rate for a star with stellar wind velocity $v$ and density $n_0$, located at a distance of $d$. We then scale all of the physical parameters in terms of Jupiter's physical parameters and note that Jupiter has an average flux density of $\Phi_{J} = 5.1\times 10^7$ Jy at 1 AU. By scaling this relation in terms of Jupiter we are assuming that Jupiter's radio emission can be described by the Radiometric Bode's law. This may not be true as the radio emission from Jupiter is not primarily driven through energy injection by the Solar wind but by the planet's rotation. Further, in using the Radiometric Bode's law we are assuming that both K2$-$33 B and WASP$-$17 b are able to convert all of the stellar wind energy flux into radio power.

The predicted flux densities for both of the known Hot Jupiter systems are listed in Table \ref{tab:HotJup}. The predicted magnetic field strength and flux density of WASP$-$17 b are not favourable for our MWA observation since the maximum frequency is less than 150 MHz and the expected flux density would be difficult to detect even with the most sensitive of today's low frequency telescopes. Note that if the magnetic field strength of WASP-17 b is best described by the scaling relations in \citet{Griessmeier:2004}, the maximum frequency would be too low to be able to detect using ground based telescopes since its below the 10 MHz ionospheric cut-off frequency. Owing to its young age and high mass, K2$-$33 b has much better predicted values for both the emission frequency and radio flux density. In section \ref{sec:last} we discuss some reasons why we did not detect any emission from this source. 

\subsubsection{Young Stars}
We do not detect any sources with fractional circular polarisation greater than 0.5 in either of the sets of stacked Stokes V images for both the \citet{Rizzuto:2011} and \citet{Rizzuto:2015} catalogues. We found 1 source in the \citet{Rizzuto:2011} catalogue to be associated with only Stokes V emission in the absolute value stacked Stokes V image on hourly time scales. We created $1 \times 1$ degree postage-stamp images for this source in Stokes I and V to further investigate this emission. We found that the source of stokes V emission is associated with leakage due to a Stokes I beam artefact from nearby bright source.

\subsection{Blind search within the field}

To carry out a blind search for exoplanet emission, again we focus solely on Stokes V emission. First we cross-matched the Stokes I and V catalogues for  each image to find sources with fractional polarisation  greater than 0.5. Once more, we did not find any sources on any time scale with fractional polarisation $> 0.5$ for either set of stacked Stokes V images. We then investigated sources in each of the images that were not associated with any of the Stokes I sources. For the absolute value stacked images we did not find any sources associated with only Stokes V emission on any time scale. For the normal stacked Stokes V images we found on hourly time scales there were 2 sources, on nightly time scales 2 sources, and in the full median stack there were no sources. Again we created $1 \times 1$ degree postage-stamp images for each of these sources to visually inspect the emission. Upon inspection of the Stokes V sources in the $1 \times 1$ degree postage-stamp images, all four of the sources were found to be associated with leakage from a beam artefact from a nearby source. 

\section{Discussion and Conclusions}\label{sec:last}
We used the MWA to observe the Upper Scorpius Association over five consecutive nights in June 2015. The purpose of these observations was to look for 154 MHz emission from young Hot Jupiter systems in this region. This is the first radio survey to target a region of young stars to look for exoplanets. We restricted our search to the Stokes V images of Upper Scorpius because the exoplanet emission is believed to be almost 100\% circularly polarised. This has the added benefit of the circularly polarised images having better sensitivity than the total intensity images. In this search we targeted known stars which could host an exoplanet system as well as two known Hot Jupiters systems, WASP-17 b and K2-33 b.  Additionally we did a blind search for sources with high fractional circular polarisation and sources identified only in our Stokes V images. Both the targeted and blind searches resulted in non-detections and we place the first limits on the radio emission from WASP-17 b and K2-33 b.

While the flux density and magnetic field strength predicted for the Solar-age WASP-17 b are too low to be detected by the MWA, the flux density and field expect for the much younger K2-33b are within the detection limits of this telescope. To determine whether known exoplanet systems are likely to lie within the flux density and frequency limits of the MWA, we use the  Extrasolar Planetary Encyclopaedia to identify 682 exoplanet systems with $a\leq$5.2\,AU and calculate the expected radio flux density and maximum observing frequency for each source using the relevant information. We find that 10 out of the 682 sources, or 1$\%$ of these objects, are within the sensitivity and frequency limits of the MWA. As noted before, the majority of these objects have ages $\geq$ 1 Gyr. Selecting sources with ages $<$1 Gyr, 6 of the remaining 43 sources, or 14$\%$, are within the sensitivity and frequency limits of the MWA. So we expect a small portion of unknown young Hot Jupiters within this field to have flux densities and field strengths within the limits of the MWA. There are several possible reason why we didn't detect any exoplanets in this field. 

The first reason is that we could have over estimated the expected flux density from Hot Jupiters in this field. Modelling efforts through out the literature provide a wide range of expected flux densities. In fact, those by \cite{Griessmeier:2011} are about an order of magnitude higher than ours. These difference are mainly due to the poorly constrained stellar wind velocity close to the host star, in a region where the wind has yet to reach its terminal velocity. \citet{Vidotto:2010} point-out that many models over estimate the expected flux density because they assume the terminal velocity of the wind. Without having better constraints on the wind parameters the order of magnitude uncertainties in the radio flux density estimates will remain.  

Further, in our modelling we assume the emission is driven by the kinetic energy Radiometric Bode's law, but its possible that the magnetic case may dominate in most cases. \citet{Griessmeier:2007} estimate flux densities on the order of 10 mJy for the situation of magnetic energy dominate planet-star interaction. This would be challenging to detect with the MWA. Even worse still, \citet{Nichols:2016} argue that due to the saturation of magnetosphereic convection for Hot Jupiters, these systems will not be able to dissipate the total incident magnetic energy. Thus the magnetic Radiometric Bode's law over estimates the flux densities substantially. \citet{Nichols:2016} calculate flux densities on the order of 1 mJy in the case of convection saturation. Future low-frequency telescopes, like the Square Kilometre Array \citep{Lazio:2009}, will have better sensitivity and could make a detection if the flux densities are an order of magnitude lower.

A second issue is that we may have over estimated the magnetic field strength and thus the maximum frequency of the radio emission. The scaling laws in \citet{Griessmeier:2004} predict magnetic field strengths much lower than those of \citet{Reiners:2009}. For example, the magnetic fields estimated for K2-33 b using \citet{Griessmeier:2004} are associated with maximum frequencies that are undetectable with the MWA. It is currently not known which relations, those in \citet{Griessmeier:2004} or \citet{Reiners:2009}, more accurately predict the magnetic field strength of a given planet. However, a radio detection would be one way to discriminate between these two models.

The last issue we would like to address is the beaming of the radio emission. Radio emission from exoplanets is most likely orbitally beamed. This is seen in other objects known to produce CMI emission, such as the Solar System planets and ultracool dwarfs. For orbital periods between $1 - 6$ days, our 16.6 hours of total observation only covers between $11 - 60\%$ of these orbital periods. Furthermore, the inclination of the planetary rotation axis and the orbital axis of the system will determine whether the beam of emission sweeps in the direction of the Earth. For example, \citet{Hess:2011} find that for inclinations of the planetary orbital plane $\geq$60 degrees no emission is detectable independent of the inclination angle of the magnetic axis. If we only expect $\sim$10 Hot Jupiters in this field it is possible that a combination of unfavourable orbital geometry and lacking full orbital coverage could have prevented us from detecting exoplanetary 154 MHz emission. To mitigate these problems, constant monitoring of a large number of young stars could ensure we observe objects with favourable geometries during orbital phases that sweep emission in the direction of the Earth. 

In the future we plan to carry out similar MWA observing campaigns of other nearby star forming regions. By monitoring large samples of young stars in these regions  we can build up a census of transit phenomena in these regions and target the best candidates to lead to a radio detection of an exoplanet. 

\section*{Acknowledgements}
DLK was supported by NSF grant AST-1412421. TM acknowledges the support of the Australian Research Council through grant FT150100099. This scientific work makes use of the Murchison Radio-astronomy Observatory, operated by CSIRO. We acknowledge the Wajarri Yamatji people as the traditional owners of the Observatory site. Support for the operation of the MWA is provided by the Australian Government (NCRIS), under a contract to Curtin University administered by Astronomy Australia Limited. We acknowledge the Pawsey Supercomputing Centre which is supported by the Western Australian and Australian Governments.This research was conducted by the Australian Research Council Centre of Excellence for All-sky Astrophysics (CAASTRO), through project number CE110001020. Special thanks to the anonymous referee whose comments greatly improved this manuscript.

%%%%%%%%%%%%%%%%%%%%%%%%%%%%%%%%%%%%%%%%%%%%%%%%%%



\bibliographystyle{mnras}
\bibliography{exoplanet} % if your bibtex file is called example.bib

%%%%%%%%%%%%%%%%%%%%%%%%%%%%%%%%%%%%%%%%%%%%%%%%%%

% Don't change these lines
\bsp	% typesetting comment
\label{lastpage}
\end{document}